\documentclass[twocolumn, nofootinbib]{revtex4-2}
\usepackage{amsfonts}
\usepackage{amssymb}
\usepackage{amsthm}
\usepackage{amsmath}
\usepackage{graphicx}
\usepackage{hyperref}
\usepackage{enumerate}
\usepackage{tikz}
\usetikzlibrary{decorations.pathmorphing, patterns,shapes}
\usepackage{color}
\usepackage[mathscr]{euscript}

\usepackage{soul,todonotes}

\usepackage{mathrsfs}
\usepackage{hyperref}
\usepackage{datetime}

\numberwithin{equation}{section}

\newcommand{\ii}{{\rm i}}
\newcommand{\ee}{{\rm e}}
\newcommand{\x}{{\rm x}}

\newcommand{\dd}{{\rm d}}

\newtheorem{thm}{Theorem}

\theoremstyle{definition}

\newdateformat{daymonthyear}{\THEDAY \, \monthname[\THEMONTH] \THEYEAR}

\newcommand{\Cf}{\mathscr{C}}

\begin{document}

\title{A new class of semiclassical gravity solutions, gravitational quantum stealths and regular Cauchy horizons}
\author{Benito A. Ju\'arez-Aubry}
\affiliation{ Department of Mathematics, University of York, Heslington, York YO10 5DD, UK \\ benito.juarezaubry@york.ac.uk}
\date{\daymonthyear\today}

\begin{abstract}
We consider semiclassical gravity with a Klein-Gordon field with mass $m^2 \geq 0$ and curvature coupling $\xi = 1/2$. We identify a special class of Hadamard two-point functions for which the semiclassical system reduces to a quasi-linear hyperbolic system and, within this class, provide the first well-posedness result for semiclassical gravity without spacetime symmetries. These two-point functions yield stress-energy expectation values proportional to the spacetime metric (modulo ambiguities). If the proportionality constant vanishes, one has a quantum version of  gravitational `stealth' configurations. We discuss that some of these solutions of this reduced semiclassical system have regular Cauchy horizons in the light of strong cosmic censorship.
\end{abstract}

\maketitle

\section{Introduction}
\label{sec:Intro}

Recent years have seen a vigorous exploration of the strong regime of gravity, by a combination of experimental observations, especially in gravity wave astronomy since \cite{LIGOScientific:2016aoc}, with analytical and numerical predictions, see e.g. \cite{Barack:2018yly}. The motivation is understanding whether general relativistic deviations can be measured in today's experiments; this could potentially inform us about quantum gravity. 

The most conservative theory superseding general relativity is semiclassical gravity, which lies presumably in the semiclassical regime of quantum gravity, below the Planck scale. This theory describes an effective coupling between quantum fields and classical gravity through the semiclassical Einstein equations, where the expectation value of the quantum stress-energy tensor sources the Einstein field equations, while the quantum fields propagate in the spacetime that they curve. 

However, due to  technical challenges, rigorous results of semiclassical gravity and examples have been obtained only in highly symmetric spacetimes\footnote{See e.g. \cite{Eltzner:2010nx, Gottschalk:2018kqt, GottRothSiem, Janssen:2022vux, Juarez-Aubry:2019jbw, Juarez-Aubry:2019jon, Juarez-Aubry:2020uim, Juarez-Aubry:2021tae, Juarez-Aubry:2021abq, Juarez-Aubry:2024slj, Meda:2020smb, Meda:2021zdw, Meda:2022hdh, Pinamonti:2010is, Pinamonti:2013wya, Pinamonti:2013zba, Sanders:2020osl}.} and mathematical statements on the full theory remain conjectures, even for a Klein-Gordon field as a matter model \cite{Juarez-Aubry:2022qdp}. 

In recent years, there has been mounting evidence that strong cosmic censorship might hold semiclassically, even if it fails classically. A case that gives hope is Reissner-Nordstr\"om-de Sitter \cite{Hollands:2019whz, Hollands:2020qpe}, where the quantum stress-energy tensor diverges more strongly than the classical one at the inner horizon.

In this letter, we obtain novel structural results in spacetimes without any symmetries, imposing instead some conditions on the quantum matter side. We show that there exist a sector of semiclassical gravity with a non-minimally coupled scalar field, for which unique solutions exist, given suitable initial data, without making assumptions on the spacetime symmetries, when the curvature coupling of a Klein-Gordon field is $\xi = 1/2$.

More precisely, we consider the system
\begin{subequations}
\label{SGeq}
\begin{align}
 G_{ab} + \Lambda g_{ab}   = \kappa \langle T_{ab} \rangle_\omega, \\
 (\Box - m^2 - \frac{1}{2}R(\x)) \langle \Phi(\x) \Phi(\x') \rangle_\omega  = 0 \nonumber \\
 = (\Box' - m^2 - \frac{1}{2}R(\x')) \langle \Phi(\x) \Phi(\x') \rangle_\omega, \label{Wightman}
\end{align}
\end{subequations}
where $\langle \cdot \rangle_\omega$ denotes the expectation value in the state $\omega$.

We shall show that there exists a class of Hadamard two-point functions for which Eq. \eqref{SGeq} reduce to a quasi-linear hyperbolic problem for which, given suitable, constrained analytic data on a Riemannian manifold, $\mathscr{C}$, there exists a unique analytic solution to the system \eqref{SGeq}, consisting of a pair, $(\mathscr{M}=(M, g_{ab}), \omega_2)$, such that $\mathscr{C}$ is a Cauchy surface of $\mathscr{M}$ and $\omega_2(\x, \x') = \langle \Phi(\x) \Phi(\x') \rangle_\omega$ is a Hadamard two-point function (perhaps up to positivity) for the quantised Klein-Gordon field, with the special feature that the state-dependent part of the expectation value of the stress-energy tensor is proportional to the spacetime metric. Indeed, it is this special feature that allows one to reduce the System \eqref{SGeq} to hyperbolic quasi-linear form.

This result, summarised in Theorem \ref{Thm:Main} below, allows one, in principle, to obtain examples of semiclassical gravity solutions in spacetimes without symmetries, possibly by numerical methods and is, to the best of our knowledge, the first structural result on semiclassical of this kind.

After some preliminaries in Sec. \ref{Sec:Preliminaries}, we introduce our special class of two-point functions in Sec. \ref{Sec:Stealth}. We show that some two-point functions give rise to expectation values for the stress-energy tensor which vanish (modulo renormalisation ambiguities). This is a quantum counterpart of the {\it gravitational stealth} property, see e.g. \cite{Ayon-Beato:2005yoq}, which has gained substantial attention in classical gravity. {\it Classical stealths} are classical field configurations, typically with non-trivial potentials, that have vanishing stress-energy tensor. The prototypical example \cite{Ayon-Beato:2005yoq} is for a classical, non-minimally coupled scalar field, $\phi$, in $n$-dimensional Minkowski spacetime with potential $V = \frac{2 \xi^2}{(1-4 \xi)^2} \left( \lambda_1 \phi^{\frac{1-2 \xi}{\xi}} + 8(n-1)(\xi - \xi_c) \lambda_2 \phi^{\frac{1}{2\xi}} \right)$, where $\lambda_1$ and $\lambda_2$ are constants, $\xi$ is the curvature coupling parameter and $\xi_c = (n-2)/(4(n-1))$ is the conformal value of the curvature coupling. Relating the parameters $\lambda_1$ and $\lambda_2$, certain solutions, including instantons, shock waves or tachyons satisfy that their stress-energy tensor vanish.

It is notable that, with a quantum field, the stealth property can be achieved with a linear scalar field.\footnote{An interesting, separate question is whether the classical stealth property is robust under quantisation, and our result does not provide insight on this point. The quantum stealths here found do not correspond to classical stealth configurations.}

We then move to Sec. \ref{Sec:Well-posed}, where details on the well-posedness of the semiclassical gravity system for this special class of states are discussed. The result follows from well-known results in of higher-derivative gravity, appearing in previous literature \cite{Noakes:1983xd, Morales:2018imi}. A precise statement of the result appears in Theorem \ref{Thm:Main}.

Finally, in Sec. \ref{Sec:Conc}, we explain how solutions to the reduced system include situations with regular (indeed analytic) Cauchy horizons, but an examination of these solutions does not immediately lead to violations of the strong cosmic censorship conjecture in the semiclassical regime of gravity \cite{Juarez-Aubry:2023kvl}. This is especially interesting, since it has been argued that semiclassical gravity can help rescue strong cosmic censorship when it fails in classical general relativity. More precisely, if the state positivity axiom is relaxed, one can construct exact solutions with regular Cauchy horizons and finite (even vanishing) stress-energy tensor at the horizon.

\section{Preliminaries}
\label{Sec:Preliminaries}

\subsection{Hadamard states}

Hadamard states constitute the class of physical quantum states in curved (and flat) spacetimes. In a fixed globally hyperbolic spacetime of four spacetime dimensions, a state for the Klein-Gordon field (with $m^2 \geq 0$ and $\xi \in \mathbb{R}$) is called Hadamard if in a normal convex neighbourhood its Wightman function takes the form
\begin{align}
 \langle \Phi(\x) \Phi(\x') \rangle_\omega  & = \lim_{\epsilon \to 0^+} H_{\ell\epsilon}(\x, \x') + \frac{1}{2(2 \pi)^2} W(\x, \x'),
\label{HadamardCondition} \\
H_{\ell \epsilon} (\x, \x') & := \frac{1}{2 (2 \pi)^2} \left[ \frac{\Delta^{1/2}(\x, \x')}{\sigma_\epsilon(\x, \x')} \right. \nonumber \\ 
& \left. +  V(\x, \x') \log \left(\frac{\sigma_\epsilon(\x, \x')}{\ell^2} \right)  \right]. \label{HadamardBiDis}
\end{align}

Here, $H_{\ell \epsilon}$ is (a regularised version of) the Hadamard parametrix of the Klein-Gordon equation at length scale $\ell$, $\Delta$ is the van Vleck--Morette determinant, $\sigma(\x, \x')$ is half the squared geodesic distance between the points $\x$ and $\x'$ and $\sigma_\epsilon(\x, \x') = \sigma(\x, \x') + 2 \ii \epsilon(t(\x) - t(\x')) + \epsilon^2$ is a regularised version thereof that prescribes the distributional singularities of the Wightman function, where $t$ is an arbitrary time function. The Hadamard coefficients $V$ and $W$ are smooth, symmetric bi-scalars, which satisfy the following equations in each argument
\begin{align}
 (\Box - m^2 - \xi R ) V & = 0, \label{VEq}\\
 \sigma(\Box - m^2 - \xi R) W & = -(\Box - m^2 - \xi R) \Delta^{1/2}- 2 V \nonumber \\
 & - 2 V_{;a} \sigma^{;a} + 2 V \Delta^{1/2} (\Delta^{1/2})_{;a} \sigma^{;a}, \label{WEq}
\end{align}
subject to some boundary conditions at coincidence points.

We refer to two-point functions of the form of Eq. \eqref{HadamardCondition} as Hadamard two-point functions. Any Hadamard two-point function defines a quasi-free state, up to positivity. Furthermore, Hadamard two-point functions differ only by their smooth bi-scalar $W$, since $V$ is determined uniquely by the geometry of spacetime and the parameters of the Klein-Gordon equation. In analytic spacetimes, the coefficients $V$ and $W$ admit convergent series \cite{Garabedian},
\begin{align}
V(\x, \x') & = \sum_{n = 0}^\infty V_n(\x, \x') \sigma^n(\x, \x'), \label{Vexp} \\
W(\x, \x') & = \sum_{n = 0}^\infty W_n(\x, \x') \sigma^n(\x, \x'),
\label{Wexp}
\end{align}
which, when fed into Eq. \eqref{VEq} and \eqref{WEq} yield the well-known Hadamard recursion relations. It is convenient to express each bi-scalar in the series \eqref{Vexp} and \eqref{Wexp} in terms of a covariant Taylor series. Using the notation of \cite{DecaniniFolacci} we write
\begin{align}
V_n(\x, \x')  & = v_n(\x) + \sum_{k = 1}^\infty \frac{(-1)^k}{k!} v_{n(k)}(\x, \x'),  \text{ with }\nonumber \\
v_{n(k)}(\x, \x')  & = v_{n a_1 \ldots a_k}(\x) \sigma^{; a_1}(\x,\x') \dots \sigma^{; a_k}(\x,\x'), \\
W_n(\x, \x')  & = w_n(\x) + \sum_{k = 1}^\infty \frac{(-1)^k}{k!} w_{n(k)}(\x, \x'), \text{ with } \nonumber \\
 w_{n(k)}(\x, \x') & = w_{n a_1 \ldots a_k}(\x) \sigma^{; a_1}(\x,\x') \dots \sigma^{; a_k}(\x,\x'). \label{Wns}
\end{align}

By the Hadamard recursion relations, in analytic spacetimes the freedom in the $W$ bi-scalar of Hadamard two-point functions can  be encoded in the symmetric bi-scalar $W_0$ or equivalently in the \emph{local} tensors $w_{0 a_1\ldots a_k}$, (cf. Eq. \eqref{Wns}) of even rank (by the symmetry of $W_0$).

\subsection{The stress-energy tensor}

The expectation value of the renormalised stress-energy tensor is defined by a point-splitting and Hadamard subtraction procedure \cite{Moretti}. Consider the operator
\begin{align}
& \mathcal{T}_{ab}   := (1-2\xi ) g_{b}\,^{b'}\nabla_a \nabla_{b'} +\left(2\xi - \frac{1}{2}\right) g_{ab}g^{cd'} \nabla_c \nabla_{d'} \nonumber \\
 & - \frac{1}{2} g_{ab} m^2 + 2\xi \Big[  - g_{a}\,^{a'} g_{b}\,^{b'} \nabla_{a'} \nabla_{b'} + g_{ab} g^{c d}\nabla_c \nabla_d  \nonumber \\
 & + \frac{1}{2}G_{ab} \Big] + \frac{1}{3} (\Box - m^2 - \xi R) g_{ab}, \label{TabPointSplit2}
\end{align}
which acts on bi-tensors. (Here, unprimed indices denote tensor indices at points $\x$ and primed indices at $\x'$.) If $\varphi$ is a classical scalar configuration, $\lim_{\x' \to \x} \mathcal{T}_{ab} \varphi(\x) \varphi(\x')$ yields the classical stress-energy tensor. The expectation value of the renormalised stress-energy tensor is defined as
\begin{align}
& \langle T_{ab}(\x) \rangle_\omega   := \lim_{\x' \to \x} \frac{1}{2 (2\pi)^2}\mathcal{T}_{ab} W (\x, \x') + \alpha_1 g_{ab}(\x) \nonumber \\ 
&  + \alpha_2 G_{ab}(\x) + \alpha_3 I_{ab}(\x) + \alpha_4 J_{ab}(\x) , \label{Stress-energyDef} 
\end{align}
where $W$ is the smooth term of the two-point function \eqref{HadamardCondition} as discussed above. 

The constants $\alpha_1 \ldots \alpha_4$ are arbitrary renormalisation ambiguities, and 
\begin{align}
& I_{ab} := R_{;ab} - \frac{1}{2} g_{ab} \Box R - \Box R_{ab} + \frac{1}{2} g_{ab} R^{cd} R_{cd} \nonumber \\
& - 2 R^{cd} R_{cadb},
 \label{Iab} \\
& J_{ab}  := 2 R_{;ab} - 2 g_{ab} \Box R + \frac{1}{2}g_{ab} R^2 - 2 R R_{ab}, \label{Jab}
\end{align}
are covariantly conserved, symmetric tensors. The first term in Eq. \eqref{Stress-energyDef} yields the state-dependent contribution to $\langle T_{ab} \rangle_\omega$.

If $m^2 = 0$ and $\xi =1/6$, the last term in Eq. \eqref{TabPointSplit2} gives rise to the trace-anomaly, as a consequence of Eq. \eqref{WEq}. Furthermore, this term vanishes classically (on-shell), while guaranteeing that $\nabla^a \langle T_{ab} \rangle_\omega = 0$ (off-shell). Using Eq. \eqref{Wns}, we can write Eq. \eqref{Stress-energyDef} in terms of the local tensors $w_0$ and $w_{0ab}$ (see the Appendix),
\begin{widetext}
\begin{align}
\langle T_{ab} \rangle_\omega & = \frac{1}{2(2 \pi)^2} \left[ - w_{0ab} + \frac{1}{4}w_{0c}{}^c g_{ab} + \frac{1}{2}  (1-2 \xi)  w_{0;ab} + \frac{1}{2} \left(2\xi - \frac{1}{2}\right)   \Box w_{0} g_{ab} 
 + \left( - \frac{ m^2}{4} g_{ab}  + \xi \left( R_{ab} -\frac{1}{4} R g_{ab} \right) \right) w_0 \right] \nonumber \\
 & + \frac{1}{4(2 \pi)^2} v_1 g_{ab} + \alpha_1 g_{ab} + \alpha_2 G_{ab} + \alpha_3 I_{ab} + \alpha_4 J_{ab},
 \label{Tabw0}
\end{align}
\end{widetext}
where
\begin{align}
v_1 & = \frac{m^4}{8} + \frac{1}{4}\left(\xi- \frac{1}{6} \right) m^2 R - \frac{1}{24} \left(\xi - \frac{1}{5} \right) \Box R \nonumber \\
& + \frac{1}{8} \left(\xi - \frac{1}{6} \right)^2 R^2 - \frac{1}{720} R_{cd} R^{cd} + \frac{1}{720} R_{abcd} R^{abcd}.
\end{align}

The conservation equation reads, in terms of $w_0$ and $w_{0ab}$ as
\begin{align}
 \langle T_{ab}{}^{;a} \rangle_\omega &  = \frac{1}{2(2 \pi)^2} \left[ - w_{0ab}{}^{;a} + \frac{1}{4}w_{0c}{}^c{}_{;b} +   \left(\frac{1}{2}- \xi\right)  w_{0;ab}{}^a \right. \nonumber \\
 & + \frac{1}{2} \left(2\xi - \frac{1}{2}\right)   (\Box w_{0})_{;b} + \left( - \frac{ m^2}{4}  -\frac{1}{4} \xi R  \right) w_{0;b}
  \nonumber \\
 &  \left.  + \xi R_{ab} w_0{}^{;a} + \frac{\xi}{4} R_{;b} w_0 \right]+ \frac{1}{4(2 \pi)^2} v_{1;b} = 0.
 \label{TabConservation}
\end{align}

That Eq. \eqref{TabConservation} vanishes is a consequence of Eq. \eqref{WEq}, which comes from the fact that the two-point function is a bi-solution to the Klein-Gordon equation.

\section{A special class of two-point functions and quantum stealths}
\label{Sec:Stealth}

We consider here two-point functions that satisfy the Klein-Gordon equation with $\xi = 1/2$, i.e. Eq. \eqref{Wightman}, and have coefficient $w_{0ab}$ given by
 \begin{align}
 \omega_{0ab} & = \frac{1}{2} R_{ab} w_0 - \gamma g_{ab},
\label{Condition}
 \end{align}
where $\gamma$ is an analytic function. By the conservation of the stress-energy tensor \eqref{TabConservation} (or Eq. \eqref{WEq}), in the case $\xi =1/2$ and using Eq. \eqref{Condition}, we have
\begin{align}
 (\Box - m^2) \omega_0 =  - 2 v_1 + 8 (2\pi)^2 \alpha_0,
\end{align}
where $\alpha_0$ is a real constant. It follows that the stress-energy tensor takes the form
\begin{align}
 \langle T_{ab} \rangle_\omega = (\alpha_0 + \alpha_1) g_{ab} + \alpha_2 G_{ab} + \alpha_3 I_{ab} + \alpha_4 J_{ab}.
\end{align}

Setting $\alpha_0 = 0$ eliminates all the state-dependent contributions to the stress-energy tensor expectation value. (Note that one can in fact set all $\alpha_i =0$ ($i = 0, \ldots 4)$ making $\langle T_{ab} \rangle_\omega = 0$ `on the nose'.) This is the quantum analogue of a classical stealth configuration \cite{Ayon-Beato:2005yoq}, as we discussed in the introduction. To the best of our knowledge, no previous quantum states with vanishing expectation value of the stress-energy tensor (up to ambiguities) had been known outside of maximally symmetric spacetimes.

\section{A quasi-linear hyperbolic system and two-point function reconstruction}
\label{Sec:Well-posed}

In this section, we address the problem of dynamically producing solutions to the semiclassical Einstein equations that will yield a globally hyperbolic spacetime with a Hadamard two-point function of the type introduced in Sec. \ref{Sec:Stealth}.

Absorbing the variables $\alpha_1$ and $\alpha_2$ into the cosmological and Newton's constant, and setting $\kappa_1 = \kappa \alpha_3$ and $\kappa_2 = \kappa \alpha_4$, the semiclassical gravity equations \eqref{SGeq} take the decoupled form
\begin{subequations}
 \label{SimplifiedEFE}
\begin{align}
  & E_{ab}  := G_{ab} + (\Lambda-\kappa \alpha_0) g_{ab} - \kappa_1\left(  R_{;ab} - \frac{1}{2} g_{ab} \Box R \right. \nonumber \\
 &\left.  - \Box R_{ab} + \frac{1}{2} g_{ab} R^{cd} R_{cd} - 2 R^{cd} R_{cdab} \right) - \kappa_2 \left( 2 R_{;ab} \right. \nonumber \\
 &  \left. - 2 g_{ab} \Box R + \frac{1}{2}g_{ab} R^2 - 2 R R_{ab} \right) = 0, \label{HigherDerivative} \\
&  (\Box - m^2) \omega_0  =  - 2 v_1 + 8 (2\pi)^2 \alpha_0. \label{Non-HomogeneousKG}
\end{align}
\end{subequations}
with the additional condition \eqref{Condition}.

Eq. \eqref{HigherDerivative} is known to form a quasi-linear hyperbolic system that is well-posed, given suitable constrained fourth order analytic (or smooth) data on an initial Cauchy surface, $\mathscr{C}$, \cite{Noakes:1983xd, Morales:2018imi} by the Cauchy-Kovalevskaya (or Leray) theorem. Meanwhile Eq. \eqref{Non-HomogeneousKG} is a non-homogenous Klein-Gordon equation for $\omega_0$, for which a unique solution exists, given suitable initial data on $\mathscr{C}$.

In order to reconstruct the two-point function in a neighbourhood of $\mathscr{C}$, we focus on the analytic case and consider a causal normal neighbourhood of $\mathscr{C}$, $\mathscr{N}$. Around each point in this causal normal neighbourhood, the bi-scalars $V$ and $W$ can be reconstructed as convergent power series with finite radius of convergence, say $N_\x \ni \x$. In the particular case of $W$, its reconstruction relies on a given $W_0$, and hence on $w_0$ and $w_{0ab}$ and on fixing all even $w_{0a_1 \ldots a_{2n}}$. All odd $w_{0a_1 \ldots a_{2n+1}}$ are obtained in terms of all the previous $w_{0a_1 \ldots a_{2n}}$ by imposing the symmetry of $W_0$, i.e., demanding that
\begin{align}
 \lim_{\x' \to \x} [\nabla_{a_1 \ldots a_{2n+1}} W_0(\x, \x') \nonumber \\
 - g_{a_1}{}^{a_1'} \cdots g_{a_{2n+1}}{}^{a_{2n+1}'}\nabla_{a_1' \ldots a_{2n+1}'} W_0(\x, \x')] = 0.
\end{align}

The main obstacle to the construction is to guarantee that the reconstructed coefficients around every point, $V_\x$ and $W_\x$, indeed give rise to smoothly defined $V$ and $W$ in the causal normal neighbourhood $\mathscr{N}$, such that $V\vert_{N_\x} = V_\x$ and $W\vert_{N_\x} = W_\x$. The key insight to overcome this issue comes from the work of Moretti \cite{Moretti:2021}, who has shown how to define the geodesic distance in domains larger than convex normal neighbourhoods (in particular in causal normal neighbourhoods of Cauchy surfaces), by refining Kay and Wald's definition of Hadamard states \cite{KayWald}. 

Provided that a spacetime is Hausdorff and paracompact, it admits an \emph{strongly convex covering}, $\mathcal{C}$, which is a covering consisting of normal convex open sets, such that if $C_1$ and $C_2$ are in the covering, whenever $C_1 \cap C_2 \neq \emptyset$, the intersection is normal convex. This guarantees that in open neighbourhoods of the form $O = \cup_{U \in \mathcal{C}} U$ there exists a smooth assignment, $\Gamma$, of geodesic segments for pairs of points in $U$, and therefore a smoothly defined $\sigma$ across $O$, which coincides with the standard $\sigma$ for pairs of points in a fixed normal convex neighbourhood. In particular, for a globally hyperbolic spacetime with Cauchy surface $\mathscr{C}$,  in a causal normal neighbourhood of $\mathscr{C}$, $\mathscr{N}$, which is subordinate to the strongly convex covering, $\mathcal{C}$, the definition of $\sigma$ is unambiguous. Eq. \eqref{VEq} and \eqref{WEq}, as well as the formal expansions of the Hadamard coefficients, now ought to be seen as subordinate to the covering $\mathcal{C}$.

Note now (see Remark 14 in \cite{Moretti:2021}) that in the causal normal neighbourhood, $\mathscr{N}$, subordinate to $\mathcal{C}$ the coefficients $V_n$ (and $\Delta$) are completely determined and unambiguously defined and $V$ is now obtained as a convergent power series in $\sigma$ with these coefficients (by the analiticity of spacetime), and $\sigma$ is now well-defined in the whole of $\mathscr{N}$. 
By construction, $V\vert_{N_\x} = V_\x$, as desired. By the same token, we can define $W$ unambiguously in terms of its power series in $\mathscr{N}$, subordinate to the strongly convex covering  $\mathcal{C}$. In this way, a (by construction Hadamard) two-point function of the form $\omega_2 = H_{\ell \epsilon} + W$, where $H_{\ell \epsilon}$ and $W$ are locally constructed by convergent Hadamard series, is unambiguously defined in  $\mathscr{N}$. Furthermore, this two-point function will satisfy Radzikowski's microlocal spectrum condition \cite{Radzikowski}.


We collect the above discussion and render it precise in the following result:

\begin{thm}
\label{Thm:Main}
 Consider the system \eqref{SimplifiedEFE} in harmonic coordinates and let $\mathscr{C} = (C, h_{ab})$ be a three-dimensional Riemannian analytic spacetime, equipped with a symmetric, analytic tensor $K_{ab}$. Suppose there are further analytic data $R_{ab}\vert_\mathscr{C} = \mathcal{R}^{0}{}_{ab}$ and $n^c\nabla_c R_{ab}\vert_\Cf = \mathcal{R}^{1}{}_{ab}$ that, together with $h_{ab}$ and $K_{ab}$, satisfy the constraint equations $n^aE_{ab} = 0$ and $n^a n^b E_{ab} = 0$. Then, there exists a unique analytic Cauchy evolution satisfying Eq. \eqref{HigherDerivative}, $\mathscr{M}=(M, g_{ab})$, such that $\mathscr{C}$ is a Cauchy surface of $\mathscr{M}$ with extrinsic curvature $K_{ab}$ and normal $n^a$ and $R_{ab}$ is the Ricci tensor of $g_{ab}$. Furthermore, given initial analytic data, $\omega_0\vert_\Cf = \Omega_0$ and $n^a \nabla_a \omega_0\vert_\Cf = \Omega_1$, there exists a unique analytic solution, $\omega_0$ to Eq. \eqref{Non-HomogeneousKG}. Given a solution $\omega_0$ and $\omega_{0ab}$ satisfying Eq. \eqref{Condition} and choosing symmetric tensors $\omega_{0a_1 \ldots a_2n}$ (for which there is some freedom), a unique Hadamard two-point function, $\omega_2$, can be reconstructed in a causal normal neighbourhood of $\mathscr{C}$, $\mathscr{N}$. The pair $(\mathscr{N}, \omega_2)$ is a solution to the semiclassical Einstein equations with a Klein-Gordon field \eqref{SGeq}. \qed
\end{thm}

\section{Examples}
\label{Sec:Eg}

\subsection{Examples of the special class of states}

We give some explicit examples of states that satisfy our hypotheses. First of all, vacuum states in maximally symmetric spacetimes satisfy our conditions. For example, in Minkowski spacetime,
\begin{align}
\omega_2^{\mathbb{M}} & = \frac{m K_1(m \sqrt{2 \sigma_\epsilon})}{4 \pi^2 \sqrt{2 \sigma_\epsilon}} = \frac{1}{8 \pi ^2 \sigma_\epsilon }  \nonumber \\
& + \frac{1}{8 \pi^2}\left( \frac{m^2 }{2} + \frac{m^4 \sigma}{8} + O(\sigma^2) \right) \log \left(\frac{m^2 \sigma_\epsilon }{2}\right) \nonumber \\
& + \frac{m^2 \left( 2 \gamma -1\right)}{16 \pi ^2}  + \frac{m^4   \left(4 \gamma -5\right)\sigma}{128 \pi ^2}+ O(\sigma^2).
\end{align}

It is easy to see that (say, at length scale $\ell^2 = 2/m^2$), $w_0 = \frac{m^2 \left( 2 \gamma -1\right)}{16 \pi ^2}$ and $w_{0ab} = 0$, which satisfies our criteria for a choice of $\gamma$ and $\alpha_0$.

For the Bunch-Davies vacuum in de Sitter spacetime with radius of curvature $\ell$,
\begin{align}
\omega_2^{\rm BD} & = \frac{\Gamma(3/2+ \nu)\Gamma(3/2 - \nu)}{ 2(2\pi)^2(2 \ell)^2} \nonumber \\
& \times F\left(3/2- \nu, 3/2+\nu; 2, \cos^2(\sqrt{\sigma/(2 \ell^2)})\right)
\end{align}
with $\nu = \sqrt{9-4(m^2 \ell^2 + 12\xi)}/2$. We have that
\begin{align}
w_0 = \frac{\left(3-12 \nu ^2\right) H_{\frac{1}{2}-\nu }+\left(3-12 \nu ^2\right) H_{\nu +\frac{1}{2}}+12 \nu ^2-1}{48 l^2}
\label{w0BD}
\end{align}
and $w_{0ab} = 0$, where $H_k$ is the Harmonic number function. In this case, again our criteria are satisfied for a choice of $\gamma$ and $\alpha_0$. The situation is similar in anti-de Sitter spacetime, see \cite{Juarez-Aubry:2024slj}.

\subsection{Semiclassical cosmology}

We now address the question of semiclassical cosmology. For simplicity, we deal only with the case $\alpha_3 = \alpha_4 = 0$. The system of equations for $g_{ab} = - \dd t_a \dd t_b + a(t) (\dd x_a \dd x_b + \dd y_a \dd y_b + \dd z_a \dd z_b )$ becomes
\begin{align}
\left( \frac{\dot a}{a} \right)^2 = \frac{\Lambda - \kappa \alpha_0}{3}, \\
2 \frac{\ddot a}{a} + \left( \frac{\dot a}{a} \right)^2 = \Lambda - \kappa \alpha_0, \\
\left(-\partial_t^2 - \frac{3 \dot a}{a} \partial_t + \frac{\Delta}{a^2}\right) w_0 = -2 v_1 + 8 (2 \pi)^2 \alpha_0, \label{w0Cosmo}
\end{align}
where $v_1$ depends on up to 4 time derivatives of $a$, together with the constraints
\begin{subequations}
\label{ConstraintsCosmo}
\begin{align}
w_{0tt} &= -\frac{3 \ddot a}{2 a} w_0 + \gamma, \\
w_{0xx} = w_{0yy} = w_{0zz} &= \frac{1}{2}(2 \dot a^2 + a \ddot a) w_0 - \gamma a^2,
\end{align}
\end{subequations}
and $w_{0 \mu \nu} = 0$ for $\mu \neq \nu$. The first two equations of the system are the Friedmann equations, with solution $a(t) = a_0 \exp [\sqrt{(\Lambda - k \alpha_0)/3}t]$. Eq. \eqref{w0Cosmo} becomes
\begin{align}
\left(-\partial_t^2 - \frac{3 \dot a}{a} \partial_t + \frac{\Delta}{a^2}\right) w_0 = -\frac{119}{270} (\Lambda -\alpha_0 \kappa )^2 + 32 \pi^2 \alpha_0\nonumber \\
+\frac{2}{3} m^2 (\alpha_0 \kappa -\Lambda ) -\frac{m^4}{4}.
\label{w0CosmoEq}
\end{align}

We can obtain explicit solutions to Eq. \eqref{w0CosmoEq}. The homogeneous problem is solved by Bunch-Davies type modes --- it suffices to transform to conformal time and the homogeneous version of Eq. \eqref{w0CosmoEq} can be cast as a Bessel equation in the time variable (see \cite[Sec. 3]{Bunch-Davies}) with plane waves in the spatial directions. A solution to Eq. \eqref{w0CosmoEq} can be obtained, for example, by adding a non-homogeneous solution of the form $C t$, where
\begin{align}
C & = \left( -3 \sqrt{(\Lambda - k \alpha_0)/3} \right)^{-1} \left[-\frac{119}{270} (\Lambda -\alpha_0 \kappa )^2 + 32 \pi^2 \alpha_0 \right. \nonumber \\
& \left. +\frac{2}{3} m^2 (\alpha_0 \kappa -\Lambda ) -\frac{m^4}{4}\right],
\end{align}
thus $w_0 = \int \dd^3 k h(k) \psi_k(t) \ee^{\ii {\bf k} \cdot {\bf x}} + C t$, where $h$ is a rapidly decaying function, solves Eq. \eqref{w0CosmoEq}. The resulting two-point function can now be in principle obtained by imposing the constraints \eqref{ConstraintsCosmo} to find $w_{0ab}$, fixing the remaining free Hadamard coefficients and using the recursion relations.

Note that the two-point function obtained by this procedure is different to the one of the usual Bunch-Davies state (cf. Eq. \eqref{w0BD}), for it is explicitly spacetime dependent, even if the stress-energy tensor is here too proportional to the spacetime metric.

\section{ Final remarks and the status of strong cosmic censorship}
\label{Sec:Conc}

The choice $\xi = 1/2$ has played a critical role in obtaining the above results, but our findings give hope that semiclassical gravity in the general case ($\xi \in \mathbb{R}$) can be well-posed, at least for a suitable class of states. Meanwhile, we have shown that local semiclassical gravity solutions exist for every (analytic) vacuum solution of higher-derivative gravity (and of general relativity!), modulo state positivity. We have also shown that every analytic curved spacetime admits a quantum stealth solution, modulo state positivity.

At this point, one can ask the question if it is possible to devise solutions to semiclassical gravity with a regular Cauchy horizon. For example, given an analytic solution of  higher-derivative gravity or general relativity ($\kappa_1 = \kappa_2 = 0$ but possibly with cosmological constant) with a Cauchy horizon, can one construct two-point functions for which Eq. \eqref{SGeq} hold, such that the expectation value of the stress-energy tensor is proportional to the spacetime metric (even vanishing) all the way up to the Cauchy horizon?

It is known since the work of Krasnikov \cite{Krasnikov} and Sushkov \cite{Sushkov:1994va, Sushkov:1995hg} that it is possible to have quantum fields on fixed backgrounds whose na\"ive stress-energy tensor remains finite at Cauchy horizons. In fact, in the work of Sushkov \cite{Sushkov:1994va} a regular quantum stress-energy tensor is constructed in a spacetime with closed timelike curves in two spacetime dimensions, which precisely has the property of being proportional to the spacetime metric (see Sec. 5 in \cite{Sushkov:1994va}), as is the case in this letter. On the other hand, these states \cite{Krasnikov, Sushkov:1994va, Sushkov:1995hg} are not Hadamard \cite{KRW} and hence the stress-energy tensor is, strictly speaking, ill defined.\footnote{Indeed, this can be seen explicitly already in \cite{Sushkov:1995hg} by computing the expectation value of the field squared.}

We proceed to discuss some new examples that have regular stress-energy tensors at a Cauchy horizon. For simplicity, we work in the `future-wedge' Kasner universe, which consists the region $t > |x|$ of Minkowski spacetime, equipped with the metric $g_{ab} = -\ee^{2 a \tau}( \dd \tau_a \dd \tau_b - \dd \chi_a \dd \chi_b) + \dd y_a \dd y_b + \dd z_a \dd z_b$, where $a > 0$, $t = a^{-1} \ee^{a \tau} \cosh(a \chi)$ and $x = a^{-1} \ee^{a \tau} \sinh(a \chi)$. Kasner's universe is globally hyperbolic in its own right and any Cauchy surface of Kasner's universe contains a past Cauchy horizon, located $t = |x|$ in Minkowski spacetime. The Minkowski vacuum is a state provides a solution to Kasner's universe in semiclassical gravity,\footnote{See \cite{Higuchi} for some details.} which is obviously regular at the past Cauchy horizon. Our prescription for the construction of two-point functions can be used to obtain new Hadamard Wightman functions (with vanishing stress-energy tensor), provided that we solve $(\Box - m^2) w_0 = m^4/8$, with $w_{0ab} = -\gamma g_{ab}$. The construction is valid in the whole domain of dependence of Kasner's universe (and beyond) because Minkowski spacetime is geodesically convex. 

Thus, it seems that we are in a position in which we can find counter-examples to strong cosmic censorship. However, these examples lie outside of the hypotheses put forth in the quantum strong cosmic censorhip conjecture \cite{Juarez-Aubry:2023kvl}, because the partial Cauchy surfaces of Kasner's universe are not `strictly partial' (see Def. 1 in \cite{Juarez-Aubry:2023kvl}). Furthermore, like in Krasnikov and Sushkov's examples, that a state has finite stress-energy tensor (as computed by some method) does not guarantee that it is Hadamard. Consider, for instance, the massless scalar field in our Kasner universe example, for which $\Box w_0 = 0$ reduces to the problem 
\begin{align}
(-\partial_\tau^2 + \partial_\chi^2) \mathscr{F}[w_0] - k_\perp^2 \ee^{-2 a \tau} \mathscr{F}[w_0] = 0, \label{KasnerEq}
\end{align} 
by Fourier transformation in the $yz$ plane. The past Cauchy horizon is approached as $t \to |x|$, i.e., as $\tau = (2 a)^{-1} \log \left((at)^2 + (ax)^2 \right) \to -\infty$, which is precisely the limit in which the potential term for $\mathscr{F}[w_0]$ in Eq. \eqref{KasnerEq} takes becomes exponentially singular. This signals a breakdown of the Hadamard property.

To finish the discussion, we emphasise that system \eqref{SimplifiedEFE} allows us for the first time to explore exact semiclassical gravity in gravitational fields without symmetries, opening the door for the analysis of important semiclassical problems exactly, such as black hole radiation and evaporation \cite{Hawking:1975}, at least in this simplified setting. We have already shown some simple application in the cosmological setting.

The issue of the positivity of the class of two-point functions here considered remains an important open question. The examples provided guarantee that this class of states is not empty. The analyticity requirement is imposed to obtain a direct argument for the reconstruction of the two-point function. The relaxation to the smooth case is left to future work.

\begin{acknowledgments}
The author is supported by the EPSRC Open Fellowship EP/Y014510/1. Warm thanks are due to E Ay\'on-Beato for discussions on stealth solutions. CJ Fewster, DW Janssen, BS Kay, P Meda and N Pinamonti are thanked for other enlightening discussions.
\end{acknowledgments}

\appendix

\section{ Derivation of Eq. \eqref{Tabw0}}

We begin by expressing the term $W = W_0 + W_1 \sigma + O(\sigma)$ appearing in the Def. \eqref{Stress-energyDef} in terms of the covariant Taylor series \eqref{Wns},
\begin{align}
\nabla_a \nabla_{b'} W & = -\frac{1}{2} w_{0;ca} \sigma^{;c}{}_{b'} + w_{0cd} \sigma^{;c}{}_{a} \sigma^{;d}{}_{b'} + w_1 \sigma_{;ab'} + O(\sigma^{1/2}) \nonumber \\
& = \frac{1}{2} w_{0;ba} g^b{}_{b'} - w_{0ab'} - w_1 g_{ab'} + O(\sigma^{1/2}), \label{Wab'} \\
\nabla_{a'} \nabla_{b'} W & = w_{0ab} \sigma^{;a}{}_{a'} \sigma^{;b}{}_{b'} + w_1 \sigma_{;a'b'} + O(\sigma^{1/2}) \nonumber \\
& = w_{0a'b'} + w_1 g_{a'b'} + O(\sigma^{1/2}), \label{Wa'b'}\\
\nabla_{a} \nabla_{b} W & = w_{0;ab} - \frac{1}{2} (w_{0;ca} \sigma^{;c}{}_{b} + w_{0;cb} \sigma^{;c}{}_{a})  \nonumber \\
& + w_{0cd} \sigma^{;c}{}_a \sigma^{;d}{}_b + w_1 \sigma_{;ab}+ O(\sigma^{1/2}) \nonumber \\
& = w_{0ab} + w_1 g_{ab}+ O(\sigma^{1/2}) \label{w1Elim}.
\end{align}

We can express the right-hand side of Eq. \eqref{w1Elim} as
\begin{align}
\nabla_{a} \nabla_{b} W & = w_{0ab} + \left(- \frac{3}{2} v_1 + \frac{1}{4}(m^2 + \xi R) w_0 -  \frac{1}{4}  w_{0c}{}^c \right) g_{ab} \nonumber \\
& + O(\sigma^{1/2}). \label{Wab}
\end{align}

To do this, one uses the boundary condition for the $V$ coefficient
\begin{align}
2 V_0 + 2V_{0;a} \sigma^{;a} - 2 V_0 \Delta^{-1/2} \Delta^{1/2}{}_{;a} \sigma^{;a} \nonumber \\
+ (\Box - m^2 - \xi R) \Delta^{1/2} = 0, 
\end{align}
to rewrite \eqref{WEq} as
\begin{align}
 (\Box - m^2 - \xi R) W & = \sigma^{-1}\left[- 2 (V-V_0) - 2 (V-V_0)_{;a} \sigma^{;a} \right. \nonumber \\
 & \left. + 2 (V-V_0) \Delta^{1/2} (\Delta^{1/2})_{;a} \sigma^{;a}\right], 
\end{align}
where the right-hand side is $O(\sigma^0)$ because $V - V_0 = O(\sigma)$. Expanding $W$ and $V-V_0$ in covariant Taylor series yields to leading order
\begin{align}
\omega_{0c}{}^c +(m^2+ \xi R) \omega_0 + 4 \omega_1 = - 6 v_1,
\end{align}
from which the relation $\omega_1 = - \frac{3}{2} v_1 + \frac{1}{4}(m^2 + \xi R) w_0 -  \frac{1}{4}  w_{0c}{}^c$ follows. Using this relation in Eq. \eqref{w1Elim} yields Eq. \eqref{Wab}. Finally, using Eq. \eqref{Wab'}, \eqref{Wab'} and \eqref{Wab} in \eqref{Stress-energyDef} and taking the coincidence limit one obtains Eq. \eqref{Tabw0}.

\end{document}